\PassOptionsToPackage{svgnames}{xcolor}
\documentclass[journal]{vgtc}

\onlineid{1121}

\vgtccategory{Research}

\title{VArify: A Visual Analytics System for Verifying Knowledge-Enhanced Large Language Model Responses in Food Science}

\author{
\authororcid{Sam Yu-Te lee}{0009-0000-2629-3954},
  Yan To Linus Lam, 
  Manami Nakagawa,
  and Kwan-Liu Ma
}
\authorfooter{%
\item
  Sam Yu-Te lee,
  Yan To Linus Lam, 
  Manami Nakagawa,
  and Kwan-Liu Ma are with the University of California, Davis.
  	E-mail: ytlee, ytllam, mnakagawa, klma@ucdavis.edu
}

\abstract{%
Graph Retrieval-Augmented Generation (GraphRAG) grounds Large Language Model (LLM) responses in structured, domain-specific knowledge. However, verification remains difficult because errors can arise at multiple layers: an answer may not be faithful to its retrieved context, the retrieval pipeline may select irrelevant evidence, and the knowledge graph may omit relevant facts or contain incorrect triplets. We present VArify, a visual analytics system\revmark{T1-1} for food-science researchers to inspect and repair these layers in a GraphRAG workflow. A formative study with five participants using a conventional chat-and-table interface revealed behaviors reflecting answer anchoring, over-trust, and difficulty tracing multi-hop relationships. In response, VArify combines an evidence-first answer reveal with coordinated chat and hierarchical graph views, provenance inspection, and subquery editing\revmark{T1-1}. It connects the generated answer, the retrieval state, and the provenance-linked graph so that researchers can trace discrepancies across them and repair the retrieved context. From a qualitative system-use study on six food science researchers, we characterize how they scrutinize GraphRAG evidence with progressive visual inspection. Our study also reveals an important verification boundary: information absent from both retrieval and a user's knowledge remains unverifiable. We discuss the broader role of visualization in supporting oversight of LLM responses and outline future research directions.

}

\renewcommand{\manuscriptnotetxt}{}

\graphicspath{{figs/}{figures/}{pictures/}{images/}{./}} 

\usepackage{tabu}                      
\usepackage{booktabs}                  
\usepackage{tabularx}
\usepackage{lipsum}                    
\usepackage{mwe}                       
\usepackage{ccicons}                   
\usepackage{soul}

\usepackage{mathptmx}                  
\usepackage{verbatim}                  
\newcommand{\system}{VArify}
\newcommand{\revmark}[1]{}

\usepackage{balance}
\usepackage{placeins}
\usepackage[most]{tcolorbox}
\definecolor{cback}{HTML}{EDEFF1}
\definecolor{cframe}{HTML}{B9C4CA}
\definecolor{cgrey}{HTML}{666666}
\definecolor{cT}{HTML}{bae6fd}
\definecolor{cF}{HTML}{DDF4FF}
\definecolor{cR}{HTML}{D8F793}
\definecolor{cMF}{HTML}{E5D9F2}

\newtcbox{\inlineDRbox}[1][]{enhanced,
 box align=base,
 nobeforeafter,
 colback=cR,
 colframe=cframe,
 size=small,
 fontupper=\footnotesize,
 left=1pt,
 right=1pt,
 boxsep=0.5pt,
 #1}

\newtcbox{\inlineFbox}[1][]{enhanced,
 box align=base,
 nobeforeafter,
 colback=cF,
 colframe=cframe,
 size=small,
 fontupper=\footnotesize,
 left=1pt,
 right=1pt,
 boxsep=0.5pt,
 #1}

\newtcbox{\inlineMFbox}[1][]{enhanced,
 box align=base,
 nobeforeafter,
 colback=cMF,
 colframe=cframe,
 size=small,
 fontupper=\footnotesize,
 left=1pt,
 right=1pt,
 boxsep=0.5pt,
 #1}

\begin{document}

\maketitle
\section{Introduction}
Graph Retrieval-Augmented Generation (GraphRAG) systems integrate Large Language Model (LLM) chatbots with knowledge-graph retrieval to synthesize domain knowledge in areas such as medicine~\cite{wu2024medicalgraphrag} and law~\cite{zhai2025law_graphrag}. Knowledge graphs represent facts as \textit{$\langle$source entity, relationship, target entity$\rangle$} triplets~\cite{ji2022survey_knowledge_graphs}; GraphRAG retrieves relevant facts and supplies them as context for response generation~\cite{edge2025graph_rag, mei2025surveycontextengineering}. Surfacing this context can provide evidence for the response, but it does not guarantee that the response or its evidence is correct.

Consequently, users cannot assess a GraphRAG response's correctness from its final answer; they must examine its relation to retrieved context, how context was selected, and whether graph facts are credible. We refer to inspecting the answer and evidence chain for errors or limitations as \emph{verification} in GraphRAG, which spans three layers. First, \emph{response grounding} asks whether claims in the generated answer are supported by retrieved context. Second, \emph{retrieval correctness} asks whether subqueries, entity mappings, and retrieved graph relationships are appropriate. Third, \emph{knowledge-graph validity} asks whether retrieved facts represent their source material and domain. \revmark{T1-2}These three categories are used to evaluate RAG systems~\cite{Yu_2025ragsurvey, salemi2024evaluatingrag}. Failures at any layer can produce a plausible but unsupported, incomplete, or incorrect response~\cite{Yu_2025ragsurvey, karpukhin2020dpr}.

Existing work addresses this through explanations, citations, uncertainty communication, and Cognitive Forcing Functions (CFFs) that encourage scrutiny of AI outputs~\cite{passi2024appropriate}. However, explanations and citations may not reveal inconsistencies between an answer and cited evidence~\cite{kim2025explanations_sources_inconsistencies}. Uncertainty cues may highlight potentially unreliable claims~\cite{Kim2024uncertainty_expression}, while CFFs introduce usability friction, e.g., delaying answers or prompting reflection before acceptance~\cite{buccinca2021cognitiveforcing, drosos2025itmakesthinkprovocations}. \revmark{T1-2}
More integrated tools may expose intermediate retrieval results from GraphRAG pipelines~\cite{wang2024ragviz, wang2025xgraphrag, li2024linkq} or support editing intermediate LLM outputs and rerunning later steps~\cite{wu2022promptchainer, arawjo2024chainforge}. Still, these approaches offer limited support for tracing discrepancies across layers or correcting retrieved context after finding a failure.

Guided by this gap, we investigate three research questions: \textbf{RQ1:} What verification breakdowns arise when users inspect GraphRAG responses and retrieved knowledge through a baseline chat-and-table interface? \textbf{RQ2:} How can a visual analytics system expose and connect the response, retrieval, and knowledge-graph layers to support progressive verification and retrieved-context repair? \textbf{RQ3:} How do food-science researchers\revmark{T2-1} verify GraphRAG responses using an evidence-first visual analytics interface?

We address these questions through a two-stage investigation using FoodAtlas, a large-scale food-science knowledge graph constructed from scientific literature~\cite{youn2024foodatlas}. First, a formative study with five participants using a baseline chat-and-table interface surfaced answer anchoring, over-trust in structured evidence, and difficulties tracing multi-hop relationships and upstream retrieval failures. \revmark{T2-1}These findings yielded design requirements for \system{}, an implemented visual analytics system\revmark{T1-2} for food-science researchers who assess food, chemical, and disease relationships against the scientific literature. It connects the response, retrieval, and knowledge-graph layers. \system{} combines an informative CFF with coordinated chat and hierarchical graph views, supporting progressive inspection from a context summary to retrieved entities, relationships, mappings, subqueries, and source excerpts. It also enables users to edit the retrieved context and regenerate the answer. Second, a qualitative system-use study examined how six participants with food-science experience applied these features. We observed pre-reveal evidence inspection, discrepancies traced into evidence and mapping decisions, and domain knowledge used to recognize graph incompleteness and a possible extraction error.

Together, the two studies illustrate how an informative CFF and visualization can direct attention toward verification, while also revealing limits imposed by knowledge-graph quality and by information absent from both retrieval and the user's knowledge. \revmark{T3-1}Our empirical observations are grounded in FoodAtlas; applying the workflow elsewhere would require inspectable retrieval state and provenance, while its graph representation would need adaptation to other structures.

Within this scope, our contributions are:\revmark{T1-2}
\begin{itemize}
    \item \system{}, an integrated system for cross-layer verification and repair of GraphRAG answers, with the design rationale behind its evidence-first workflow and coordinated views, including requirements elicited through the formative study;
    \item Qualitative observations of food-science researchers using \system{} to trace cross-layer discrepancies and recognize knowledge-graph limitations.
\end{itemize}
\section{Related Work}
We situate our work in context engineering and GraphRAG, overreliance and cognitive bias, and visual analytics for knowledge graphs. These areas frame three verification layers, human factors, and graph-inspection tasks in GraphRAG verification.

\subsection{Context Engineering and GraphRAG for LLM Chatbots}
Context engineering~\cite{mei2025surveycontextengineering} supplies LLMs with task-relevant information, and retrieval-augmented generation (RAG)~\cite{lewis2020rag, karpukhin2020dpr, fanSurveyRAGMeeting2024} improves factual accuracy in knowledge-intensive tasks by retrieving external information as model context. It provides access to external or updated knowledge without retraining and, when surfaced to users, evidence about response generation~\cite{Yu_2025ragsurvey}. GraphRAG~\cite{edge2025graph_rag} retrieves from knowledge graphs, with reported advantages over text-based RAG for structured reasoning and domain question answering~\cite{pengGraphRetrievalAugmentedGeneration2024}. 

GraphRAG remains vulnerable at each of the three verification layers: response grounding, retrieval correctness, and knowledge-graph validity. At the response-grounding layer, a model may hallucinate, misinterpret retrieved triplets, or use inappropriate assumptions despite appropriate context~\cite{huang2025LLMhallucinationsurvey}. At the retrieval-correctness layer, context may be irrelevant, incomplete, or contradictory~\cite{karpukhin2020dpr, salemi2024evaluatingrag, Yu_2025ragsurvey}; at the knowledge-graph-validity layer, automatically extracted facts may already be incomplete or incorrect~\cite{youn2024foodatlas}. These limitations motivate tracing how failures propagate into responses.

\revmark{T1-3}Interactive tools expose RAG pipelines for inspection. RAGViz~\cite{wang2024ragviz} shows answer-to-snippet attention for text RAG and supports document exclusion and regeneration. RAGTrace~\cite{cheng2025ragtrace} and RAGExplorer~\cite{tian2026ragexplorer} diagnose retrieval and generation across text-chunk RAG configurations using chunk-level evidence. XGraphRAG~\cite{wang2025xgraphrag} helps GraphRAG developers trace failing recalls through pipeline and hierarchical topic views, inspect source chunks for extracted entities and relationships, and filter evidence. In this space, \system{} brings inspection to domain-expert end users by connecting the answer, retrieval state, and provenance-linked graph while supporting correction and regeneration.


\subsection{AI Overreliance, Cognitive Bias, and Verification}
Evidence availability does not ensure scrutiny. AI overreliance occurs when users accept or act on incorrect output~\cite{passi2025overreliance}; associated factors include AI literacy, domain expertise, and task familiarity~\cite{DEC_AI_Literacy, passi2024appropriate}. These risks motivate interfaces promoting deliberate verification~\cite{danryDeceptiveExplanationsLarge2025, lee2025genai_critical_thinking}.

Prior strategies~\cite{passi2024appropriate} provide explanations and sources~\cite{kim2025explanations_sources_inconsistencies}, convey uncertainty through highlights or language~\cite{Kim2024uncertainty_expression, vasconcelos2025generation_probabilities}, or introduce Cognitive Forcing Functions (CFFs) that interrupt workflows before users accept recommendations~\cite{buccinca2021cognitiveforcing}. CFFs may delay answers, require initial judgments, or prompt consideration of risks and alternatives~\cite{drosos2025itmakesthinkprovocations}.
Explanations and citations can impose extensive source review~\cite{mertsiotaki2025interface_for_eval_llm}, while CFFs encourage reflection without locating GraphRAG failures. 

Visual analytics offer a different approach by externalizing evidence, augmenting judgment, and shaping attention and decisions~\cite{heer2019sharedrepresentation, lee2025vizcopilot}. Still, when not designed properly, visual salience and encoding can support or hinder decisions~\cite{padilla2018decision_visualizations} or produce cognitive biases~\cite{dimara2020bias_taxonomy, wall2017warning_bias}. Li et al.~\cite{liRoleVisualizationLLMAssisted2025} found that even knowledge-graph experts could over-trust incorrect LLM outputs accompanied by helpful-looking visualizations. This motivates interfaces that support discrepancy tracing without lending visualized evidence unearned authority. \system{} combines CFFs and positions visualized evidence as an object to inspect and repair, beyond presenting evidence accompanying the generated answer.


\subsection{Visual Analytics Systems for Knowledge Graphs}
Addressing these human-factors concerns requires inspectable graph evidence. Knowledge graphs represent entities and relationships for applications including domain-specific question answering~\cite{Jiang2026hypochainer}, but their scale and structural complexity impede understanding~\cite{lissandriniKnowledgeGraph2022, nararatwong2020kg_visualization}.
Graph-visualization taxonomies provide a design vocabulary for this complexity. Lee et al.~\cite{lee2006graph_task_taxonomy} classify topology, attribute, browsing, and overview tasks; Saket et al.~\cite{saket2014group_taxonomy} extend these to grouped graphs; and Filipov et al.~\cite{filipov2023network_roadmap} consolidate tasks across topology, analytic activity, and network facets. For GraphRAG verification, these categories support locating mapped entities, inspecting attributes and relationships, tracing paths, and comparing semantic groups.

Knowledge-graph systems provide visual query interfaces~\cite{Ahmet2018OptiqueVQS, vargas2019RDF_explorer} and query-result representations~\cite{deagen2022fair, wei2020vision_kg}. Practitioner interviews identified demand for LLM-assisted queries and annotated visualizations alongside concerns about hallucination and lost semantic intent~\cite{liPreliminaryRoadmapLLMs2024}. LinkQ~\cite{li2024linkq} converts natural-language questions into graph queries and node-link visualizations.
\revmark{T1-4}Prompt-chain editors such as PromptChainer~\cite{wu2022promptchainer}, AI Chains~\cite{wu2022aichains}, and ChainForge~\cite{arawjo2024chainforge} let users inspect and edit intermediate LLM outputs and rerun later steps.
Hsuan et al.~\cite{Hsuan2024KGScope} position semantically related nodes using knowledge-graph embeddings. Tu et al.~\cite{TU2024KG_PRE_VIEW} evaluated scatterplots, node-link networks, timelines, and word clouds across knowledge-graph exploration steps. \revmark{T1-4}Building on graph representations and editable intermediate state, \system{} integrates answer-to-entity linkage, hierarchical multi-hop relationship inspection, and FoodAtlas provenance. These capabilities support food-science researchers' verification and repair across response grounding, retrieval correctness, and knowledge-graph validity.
\begin{figure*}[t]
  \includegraphics[width=1.0\linewidth]{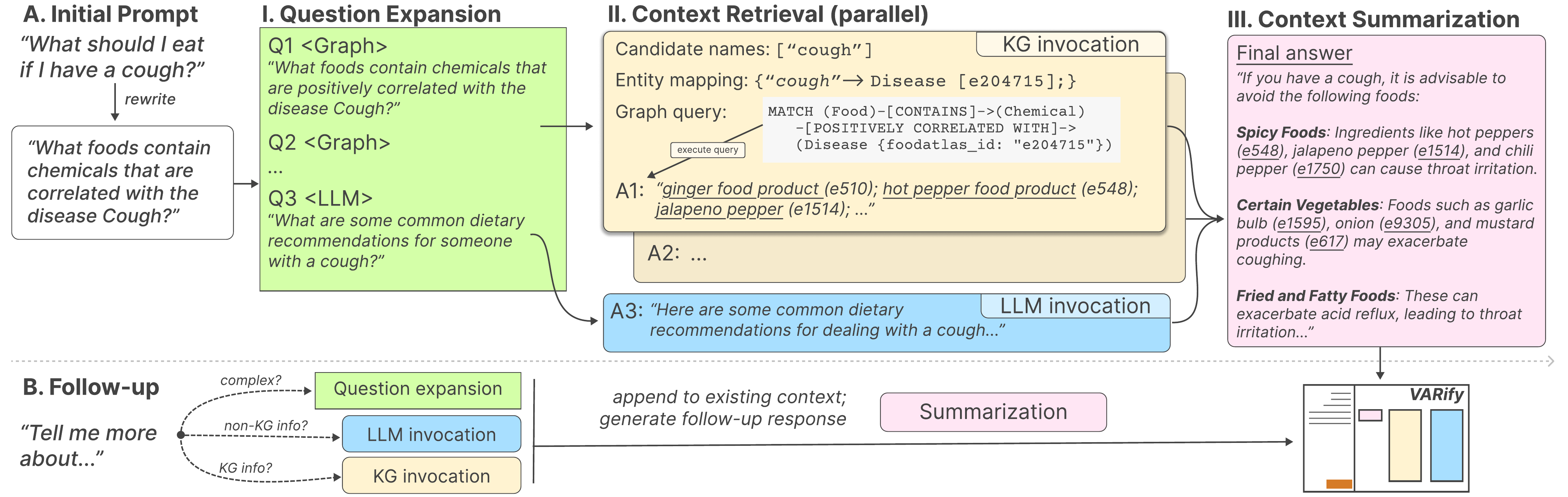}
  \caption{Data flow through the model-agnostic GraphRAG backend. (A) The prompt is rewritten using the graph schema. In I, question expansion produces subquestions routed to the graph or a standalone LLM. In II, subquestions execute in parallel. Graph retrieval comprises candidate extraction, entity mapping, query generation, and execution; standalone LLM calls receive no graph context. In III, the original prompt, subquestions, and returned answers are combined to synthesize the response. (B) Follow-ups are routed to expansion or a single graph/LLM invocation; new results are appended to the accumulated context before synthesis.}
  \label{fig:query_expansion}
  \vspace*{-0.5cm}
\end{figure*}
\section{Backend Architecture}
\label{sec:backend}
To investigate verification challenges in GraphRAG systems, we developed a four-stage pipeline consisting of prompt rewriting, question expansion, knowledge-graph context retrieval, and context summarization~\cite{mei2025surveycontextengineering}. This architecture allows us to examine verification challenges arising across common GraphRAG operations.

\subsection{Knowledge Graph: FoodAtlas}
Our GraphRAG system retrieves context from FoodAtlas, a food science knowledge graph constructed by mining nodes and relationships from over 155,000 scientific papers~\cite{youn2024foodatlas}. 
FoodAtlas contains 10,270 \textit{Foods}, 193,763 \textit{Chemicals}, and 3,177 \textit{Diseases} as entities, along with \textit{contains} relationships between food and chemical (48,474), and \textit{positively correlated with} and \textit{negatively correlated with} relationships between chemicals and diseases (89,114 positive and 49,678 negative relationships). We chose FoodAtlas as it is a large-scale food science knowledge graph that includes multi-hop relationships between entity types, e.g., \textit{oat bran} (food) contains \textit{linoleic acid} (chemical), which is \textit{negatively correlated} with \textit{hypercholesterolemia} (disease). This scale and multi-hop structure make FoodAtlas a demanding setting for examining GraphRAG retrieval and verification.
FoodAtlas records source identifiers and, for literature-derived relationships, supporting sentence excerpts as provenance. These records are imported with the knowledge graph and hosted by the backend. \system{} only retrieves and presents the records supplied by FoodAtlas.


\subsection{GraphRAG Pipeline}
\label{subsec:graphrag_pipeline}

Our GraphRAG pipeline performs four operations: prompt rewriting, question expansion, context retrieval, and context summarization, as shown in~\autoref{fig:query_expansion}. Each LLM role described below is a separate invocation. Context is shared between invocations only when explicitly passed through the pipeline state; using the same underlying model does not give an invocation access to another invocation's prompt or output. The pipeline is therefore model-agnostic and does not depend on a specific underlying LLM.

\paragraph{Prompt Rewriting}
The input to this operation is the user's original prompt. A rewriting invocation receives the prompt together with a description of the FoodAtlas schema and produces a graph-aligned version of the question. \revmark{T2-2}For example, it may rewrite ``\textit{Is there evidence that honey helps with cough?}'' as ``\textit{Which chemicals in honey are negatively correlated with cough?}'' The original prompt is retained for final response synthesis. A failure at this operation may alter the user's intent or constrain the question inappropriately to the graph schema.

\paragraph{Question Expansion}
The question-expansion invocation receives the rewritten prompt and generates four to six subquestions intended to retrieve complementary information. Its output is a set of pairs, each containing a subquestion and a destination label: \textit{Graph} or \textit{LLM}. No evidence is retrieved during this operation. Possible failures include omitting a relevant subquestion, producing duplicate or irrelevant subquestions, or routing a subquestion to an inappropriate source.

\paragraph{Context Retrieval}
The subquestions are executed independently and in parallel through one of two paths.

A standalone invocation answers an \textit{LLM}-routed subquestion without the FoodAtlas schema or graph context, using the model's parametric knowledge. This path may introduce information inaccurate or unsupported by graph evidence.

A \textit{Graph}-routed subquestion passes through components. First, an entity-extraction invocation identifies candidate food, chemical, and disease names in the subquestion. \revmark{T3-2}Second, a deterministic similarity search matches each candidate name against a full-text index of FoodAtlas entity names of the extracted type using fuzzy string matching. It retains the ten highest-scoring candidates and selects the top match without a score threshold; the others remain available for later remapping. Third, a query-generation invocation receives the subquestion, mapped entities, schema, and few-shot examples, and produces a graph query. Finally, the query returns matching FoodAtlas entities and relationships. These components may miss or misclassify an entity, select an inappropriate mapping, generate a query that misrepresents the question, or return incomplete evidence because relevant information is absent from the graph.

For both paths, the returned answer is stored with its originating subquestion and destination. Intermediate graph artifacts, including candidate entities, entity mappings, subquery destinations, and retrieved graph results, are retained separately so that \system{} can expose the retrieval process for inspection.

\paragraph{Context Summarization}
The summarization invocation receives the user's original prompt, expanded subquestions with destinations, and answers returned for each. It is instructed to synthesize the final response using only this combined answer context. The resulting response may nevertheless omit retrieved evidence, misinterpret a graph relationship, conflate graph-sourced and LLM-sourced information, or introduce a claim that is unsupported by the supplied context.

\paragraph{Follow-Up Questions}
As illustrated in~\autoref{fig:query_expansion}-B, a follow-up prompt is first rewritten using the conversation history. A routing invocation then determines whether the follow-up requires another question expansion, one graph invocation, or one standalone LLM invocation. Newly retrieved information is appended to the accumulated answer context, after which the context-summarization operation produces the follow-up response. An incorrect routing decision may retrieve insufficient or inappropriate information for the follow-up.

The pipeline was implemented using LangGraph~\cite{LangGraph}; its implementation and prompts are available in the released \system{} code repository
\footnote{\url{https://github.com/linuslyt/VArify}}. This pipeline powered the baseline interface used in the formative study described next.
\begin{figure*}[t]
  \includegraphics[width=0.95\linewidth]{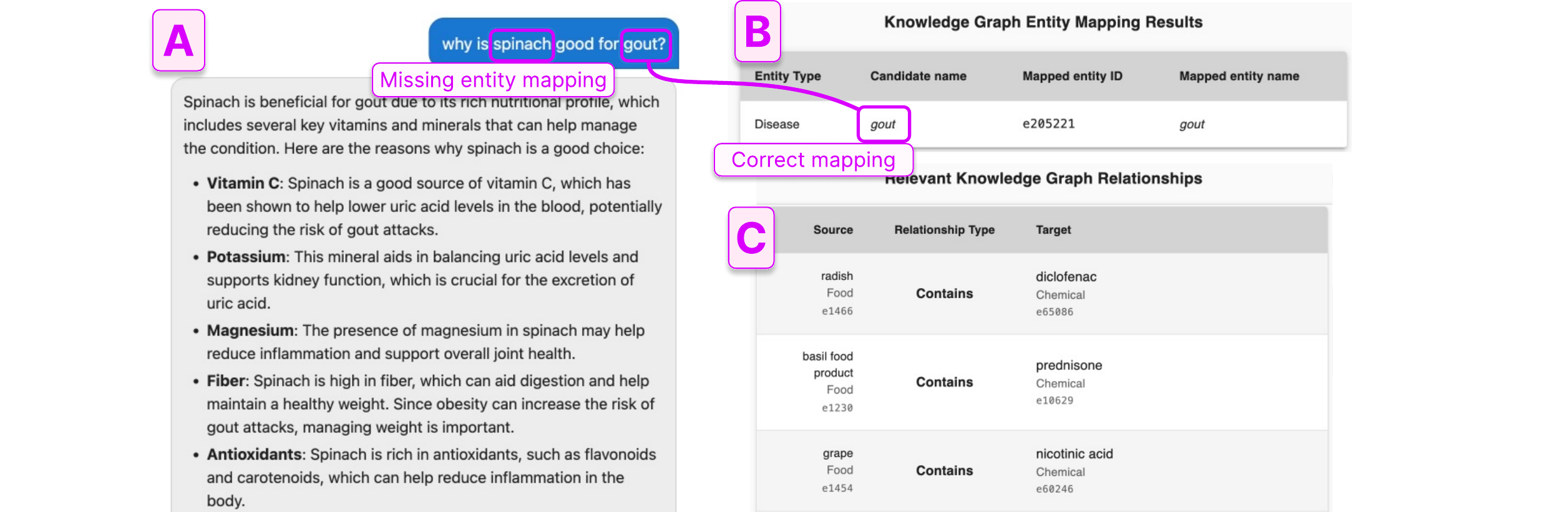}
  \caption{Baseline frontend interface used for the formative study.\ [A]: Chat Box; [B]: Entity Mapping Table; [C]: Relationship Triplet Table. In this instance, the entity mapping for \textit{Disease} \emph{``gout''} was correct, as seen by the row with matching \textbf{Candidate name} and \textbf{Mapped entity name} \emph{``gout''}. However, \emph{``spinach''} was not properly recognized as a candidate \textit{Food}, as seen by the lack of a row for \emph{``spinach''} in the entity mapping table.}
  \label{fig:pilot_interface}
  \vspace*{-0.5cm}
\end{figure*}
\section{Formative Study}
Although verification is important in GraphRAG chatbots~\cite{passi2024appropriate}, it remains unclear what interface support users need to examine answers and retrieved graph evidence together.
\revmark{T2-3}We conducted an exploratory formative requirements-elicitation study with two food-science graduate students and three computer-science graduate students to derive design requirements for interface support from the verification breakdowns we observed.\revmark{T1-5}
\revmark{T2-3}Because the resulting requirements concern interface mechanics rather than domain judgment, and because the sample was small and mixed, we treat them as design hypotheses and examine them with food-science researchers in \autoref{sec:user_study}.


\vspace*{-0.1cm}
\subsection{Baseline GraphRAG Chatbot Interface}
The baseline interface consists of a typical chat panel on the left (\autoref{fig:pilot_interface}-A), and an entity-mapping table (\autoref{fig:pilot_interface}-B) and relationship-triplet table (\autoref{fig:pilot_interface}-C) on the right.
The chat panel is integrated with the GraphRAG pipeline introduced in~\autoref{sec:backend} and supplies FoodAtlas context for food-science answers. Intermediate mappings and retrieved triplets populate their respective tables. We used tables because prior work found that users may prefer them to node-link diagrams when inspecting LLM context~\cite{liRoleVisualizationLLMAssisted2025}.
\revmark{T4-1}The interface served as a requirements-elicitation probe, so we kept it minimal and exposed exactly the three artifacts a user needs to verify an answer: the answer, the entity mappings, and the retrieved triplets.

\vspace*{-0.1cm}
\paragraph{Entity-Mapping Table}
The table displays all entities extracted from the user's query and the retrieved context (\autoref{fig:pilot_interface}-B), along with their original mentions and the mapped entity name.
For example, when the user asks \emph{``why is spinach good for gout?''}, the system recognizes that \emph{gout} is a possible name for a disease, and using similarity search, correctly matches it with the entity with FoodAtlas ID \textit{e205221} and common name \emph{gout}.
However, \emph{spinach} is missing from the entity mapping table, indicating that the system did not recognize it as an entity to search for in the knowledge graph. Its absence exposes an upstream retrieval failure that the user must detect.

\vspace*{-0.1cm}
\paragraph{Relationship-Triplet Table}
The table shows retrieved relationship triplets linked to the identified entities (\autoref{fig:pilot_interface}-C), supporting inspection of whether the LLM context was relevant. The columns \textit{Source, Relationship Type, Target} correspond to the triplet structure of retrieved facts. Rows are sorted by relationship type for readability because the retrieval pipeline provides no reliable indicator of importance. The table combines all unique triplets retrieved across GraphRAG subqueries and can therefore become a long vertical list.

\subsection{Formative Study Design}
To operationalize \textbf{RQ1}, the formative study considers three subquestions: \textbf{RQ1a:} \textit{What behaviors indicate insufficient verification of GraphRAG answers?} \textbf{RQ1b:} \textit{How do users engage with different interface components during verification?} \textbf{RQ1c:} \textit{What representational or interaction breakdowns impede verification?}
\revmark{T2-4}To elicit verification behavior, we designed quiz-based tasks that present food and disease claims for participants to assess.
For example, one task asked, ``\textit{What food is bad for edema?}''
Participants were instructed to obtain an answer from the chatbot and verify it using the baseline interface.

\paragraph{Controlled Error Injection}
\label{par:error_injection}
To ensure concrete opportunities for verification, the study interface included an experimenter-controlled error mode. The mode altered prompts and intermediate pipeline selections rather than FoodAtlas or its provenance records. During entity mapping, it randomly selected among the three highest-scoring candidates rather than always selecting the top match. It also reversed the meanings of positive and negative disease correlations in prompts used for question expansion, graph-query generation, and response synthesis. During question expansion, it requested six to eight graph-heavy subquestions while removing the normal diversity and non-duplication constraints, which could produce redundant or irrelevant context. Finally, it randomly permuted the subquestion identifiers paired with answers before response synthesis. The selected mappings, relationship labels, and retrieved triplets remained visible in the baseline interface.

The injection configuration was fixed, but its manifestations were not deterministic because mappings and subquestion identifiers were sampled and subquestions and responses were generated live; the study therefore did not prescribe a fixed number of visible errors per task. Planted errors could also coexist with unplanned pipeline failures, such as a missing candidate entity (\autoref{fig:pilot_interface}), and with inaccuracies already present in FoodAtlas. We distinguish these as \emph{planted errors}, \emph{natural pipeline errors}, and \emph{knowledge-graph errors}, respectively. Accordingly, the findings characterize how participants responded to observable discrepancies under this controlled configuration.


\vspace*{-0.1cm}
\paragraph{Participants} 
\revmark{T2-5}We recruited five participants (1 male, 4 female, aged 25--30) through research networks: two food-science graduate students (P2, P4) and three computer-science graduate students (P1, P3, P5).
\revmark{T4-2}In the findings, FS marks a food-science participant and CS a computer-science participant.
This mixed sample served requirements elicitation. Food science students contributed domain perspectives on evidence plausibility and organization, while computer science students contributed familiarity with graph and pipeline concepts. Together, these perspectives broadened the interface breakdowns considered.
Participants reported their familiarity and experience using 7-point Likert scales.
Food science students reported high familiarity with food science ($M = 6.0$, $SD=0$) and substantial experience ($M = 5.5$ years).
Computer science students reported moderate familiarity with graph-related concepts and technologies ($M = 5.3$, $SD=1.15$).
All participants were generally familiar with LLM-based chatbots ($M = 5.6$, $SD=0.8$) with reported daily usage.

\vspace*{-0.1cm}
\paragraph{Procedure}
The formative study protocol was approved by the institutional review board.
The study was conducted in person and lasted approximately one hour per session.
We first provided an overview of the study, including an introduction to knowledge graphs, the FoodAtlas dataset, and our baseline interface.
Participants were told that responses might contain AI errors; the instruction did not guarantee an error in any particular task or identify its type or location.
After providing verbal consent, participants completed a pre-questionnaire that collected information on their demographics and background knowledge.

Participants were given three quizzes via a Google Form.
For each quiz, participants interacted with our prototype to collect and verify information while composing their answers. 
They were instructed to think aloud during this process, verbalizing their reasoning and interactions. 
After submitting their answer, they completed a post-task questionnaire before proceeding to the next quiz.
Finally, participants completed a semi-structured interview (10–15 minutes) on user experience and reflections on verification.
All sessions were recorded using screen and audio capture via Zoom, and system logs were collected. 
Participants received \$20 USD in cash as compensation.



\vspace*{-0.1cm}
\subsection{Findings}
Across tasks, participants often interpreted answers before inspecting evidence, relied on surface-level matches, and struggled to reconstruct multi-hop relationships. \revmark{T4-3}One author analyzed the think-aloud sessions and interviews inductively without a predefined codebook, yielding \inlineFbox{F1}--\inlineFbox{F3}, which address insufficient-verification behaviors (\textbf{RQ1a}), use of interface components (\textbf{RQ1b}), and representational breakdowns (\textbf{RQ1c}). \revmark{T4-3}Answer-first and limited inspection appeared in both participant groups; food-science participants additionally contributed plausibility judgments and the grouping suggestion reported in \inlineFbox{F3}.

\noindent{\inlineFbox{F1}: \textbf{Behavior consistent with answer anchoring and confirmation seeking} (\textbf{RQ1a}, \textbf{RQ1b}).}
Participants began with the generated answer and searched the tables for matching evidence rather than constructing an answer from the evidence. \revmark{T4-3}P1 (CS), P4 (FS), and P5 (CS) explicitly described reading the chat response first; after forming an initial interpretation, P5 (CS) used the tables to locate elements that supported it. Expectations also affected when inspection stopped: P2 (FS) skipped verification when an output matched expectations, while P3 (CS) accepted responses that ``looked generic enough.'' Because the protocol does not establish participants' internal cognitive states, we characterize these observations as consistent with answer anchoring and confirmation seeking.

\noindent{\inlineFbox{F2}: \textbf{Behavior consistent with over-trust in structured evidence} (\textbf{RQ1a}, \textbf{RQ1b}).}
Participants sometimes accepted an answer after checking only a few entries in the relationship table. \revmark{T4-3}P1 (CS) explained that after verifying ``a couple of strings of information,'' the answer ``looked right,'' while P2 (FS) relied on the table because ``that's all we have to believe.'' Participants often searched for terms from the answer rather than reconstructing multi-hop relationships, particularly when responses were long. This limited inspection is consistent with over-trust in structured evidence and with prior findings that visual representations can unintentionally increase reliance on AI outputs~\cite{liRoleVisualizationLLMAssisted2025, liKnowledgeGraphsPractice2024}.

\noindent{\inlineFbox{F3}: \textbf{Tables impeded structural interpretation} (\textbf{RQ1b}, \textbf{RQ1c}).}
Participants had difficulty tracing a relationship discrepancy upstream to an entity mapping and reconstructing food--chemical--disease paths distributed across table rows. They also interpreted repeated entities as more important and row order as an indication of relevance, although neither encoding represented importance. \revmark{T4-3}Participants requested separate views of entities and relationships, and P4 (FS) suggested organizing evidence into broader food or chemical groups. These observations indicate that the tables did not adequately communicate the structure or hierarchy of the retrieved graph.

\vspace*{-0.1cm}
\subsection{Design Requirements}
\label{subsec:drs}
Together, \inlineFbox{F1}--\inlineFbox{F3} motivate three design requirements (\inlineDRbox{DRs}) for evidence-first interaction, structural interpretation of retrieved knowledge, and response--evidence alignment in food-science GraphRAG:

\noindent{\inlineDRbox{DR1}: \textbf{Encourage evidence-first and deliberate verification.}}
\label{par:dr1}
The answer-first and limited-inspection behaviors in \inlineFbox{F1} and \inlineFbox{F2} motivate presenting evidence before the generated response and prompting users to inspect it before accepting the answer.
One approach is to incorporate cognitive forcing functions (CFFs)~\cite{buccinca2021cognitiveforcing}, such as delaying answer reveal, requiring user interaction before showing the response, or prompting users to verify specific claims. 
These mechanisms can interrupt habitual reliance on the first response and encourage more deliberate verification.
Because trivial interruptions can disrupt workflows and eventually be bypassed, the CFF should itself convey verification-relevant information and fit naturally into verification workflows.

\noindent{\inlineDRbox{DR2}: \textbf{Support structured and domain-aligned reasoning.}}
\label{par:dr2}
The surface-level matching in \inlineFbox{F2} and difficulty reconstructing multi-hop relationships in \inlineFbox{F3} motivate a representation aligned with the graph structure and food-science concepts.
Representations should make multi-hop relationships and domain hierarchies explicit. To support this reasoning without overwhelming users, they should combine retrieved triplets with relevant domain metadata, such as ontologies, external databases, and source literature, through progressive disclosure.

\noindent{\inlineDRbox{DR3}: \textbf{Support tracing and correcting response--evidence alignment.}}
\label{par:dr3}
The answer-first and surface-matching strategies in \inlineFbox{F1} and \inlineFbox{F2} made response--evidence alignment manual and shallow, while \inlineFbox{F3} shows the difficulty of tracing discrepancies through multi-hop relationships and upstream mappings. Interfaces should connect response claims to relevant graph entities and relationships while exposing retrieval artifacts such as subquestions and entity mappings. They should also allow users to correct retrieved context and regenerate the answer, enabling verification to inform repair rather than ending at inspection.
\begin{figure*}[t]
  \includegraphics[width=1.0\linewidth]{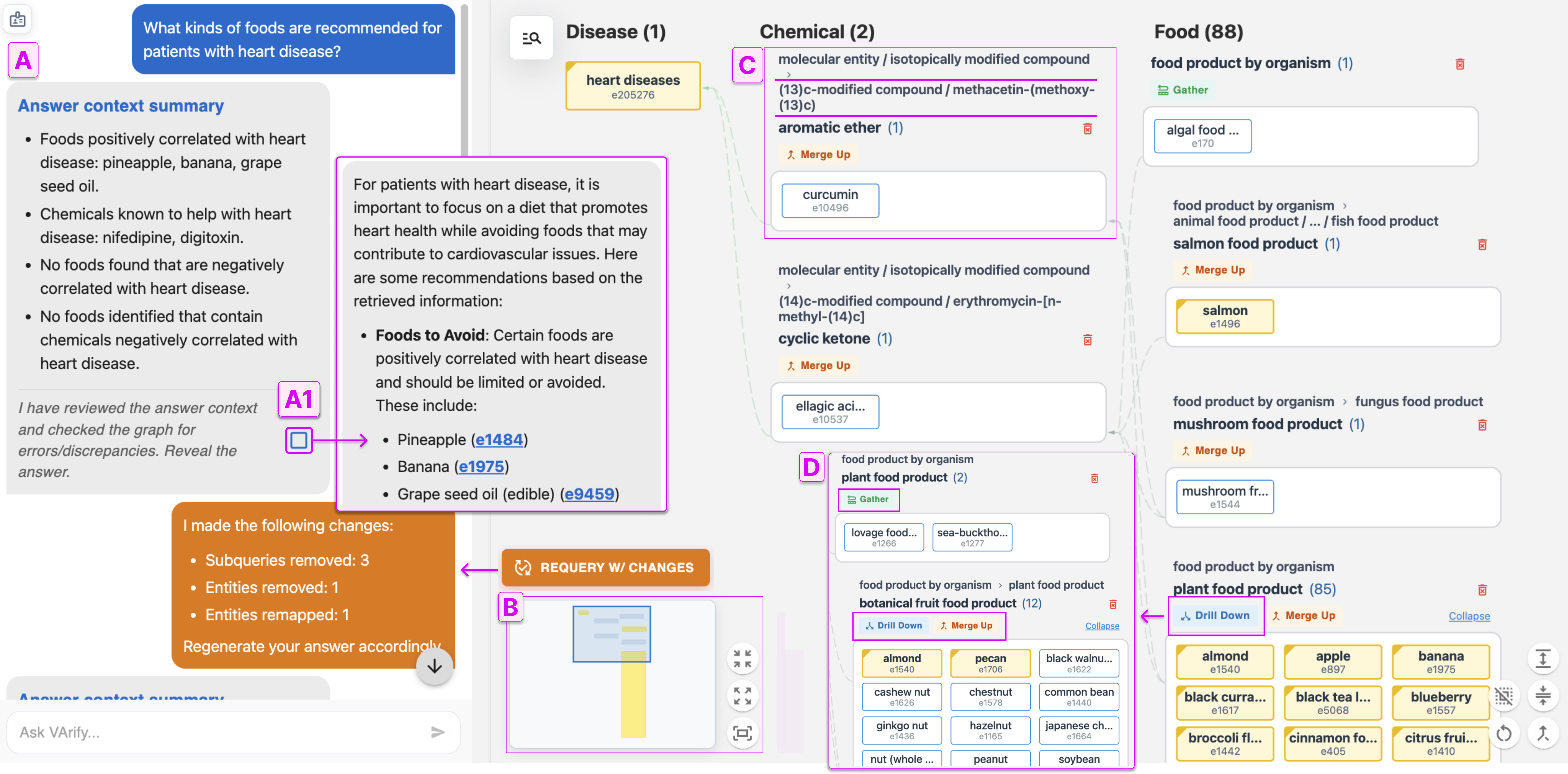}
  \caption{The interface of \system{}. [A]: chat interface with the answer context summary. Clicking the checkbox reveals the final LLM response (A1); [B]: a focus+context minimap for the graph panel. After the user edits the context, the ``Requery With Changes'' button regenerates the summary and answer using the updated retrieval state. [C]: The graph panel shows retrieved nodes and relationships in a hybrid hierarchical node-link representation. Breadcrumbs and indentations indicate hierarchy levels. [D]: Hierarchy controls support on-demand exploration.}
  \label{fig:final_interface}
  \vspace*{-0.3cm}
\end{figure*}

\begin{figure*}[t]
  \includegraphics[width=1.0\linewidth]{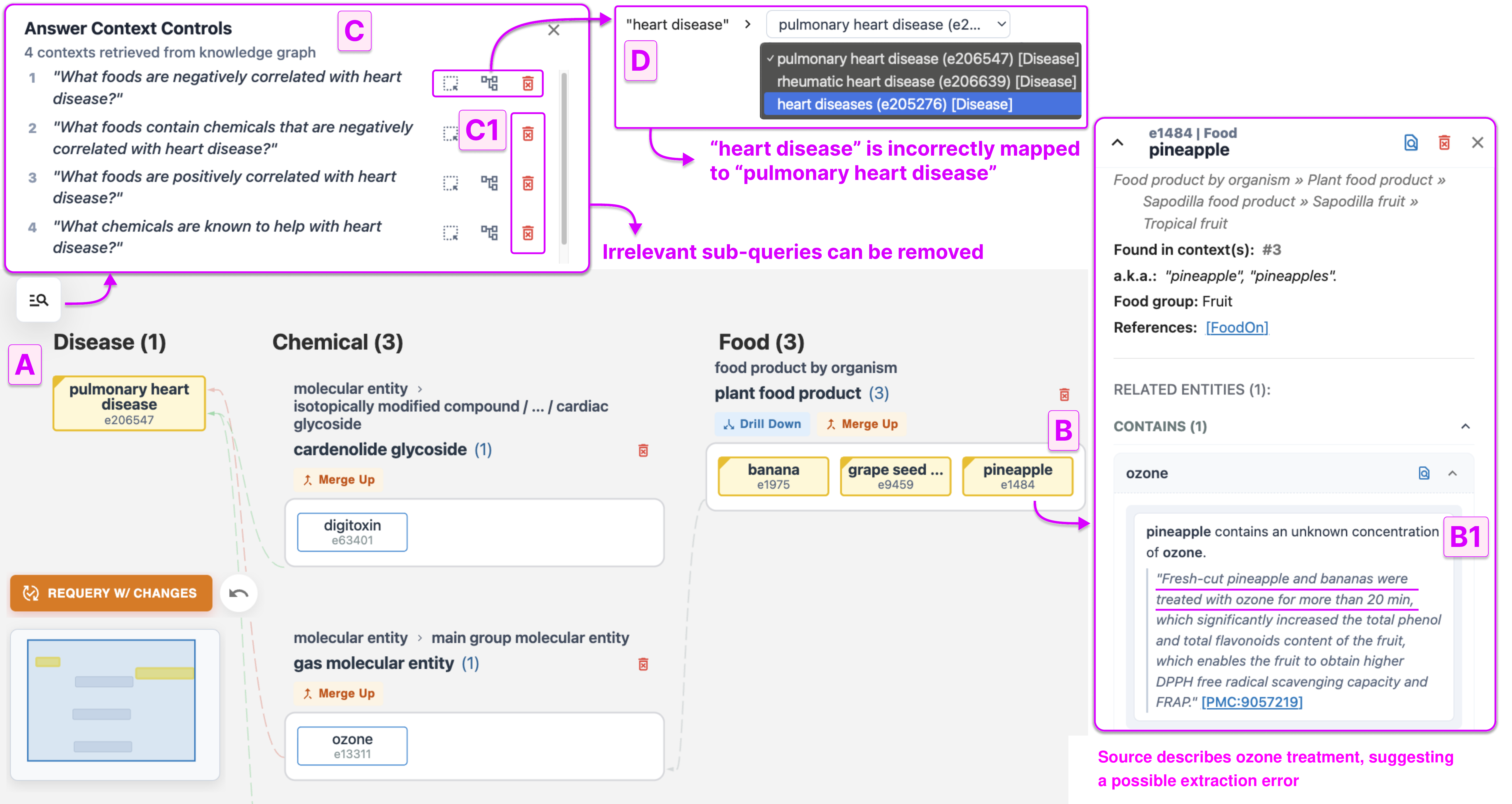}
  \caption{Verification interactions in the illustrative scenario. [A] Alice locates ``pineapple'' in the graph, linked to heart disease through ozone. [B] Alice inspects pineapple metadata and finds the source describes ozone treatment rather than natural composition, suggesting an extraction error. [C] Alice opens the answer context control panel to inspect why so few chemicals and foods are relevant to heart disease. [C1] Alice finds irrelevant subquestions and removes them. [D] Alice finds that ``heart disease'' is incorrectly mapped to ``pulmonary heart disease,'' causing the issue.}
  \label{fig:case_study}
  \vspace*{-0.5cm}
\end{figure*}
\section{Overview of \system{}}
Informed by the design requirements (DRs), we developed \system{}, a visual analytics system\revmark{T1-6} with which food-science researchers inspect and correct the evidence behind a GraphRAG answer by connecting the generated answer, the retrieval state, and the retrieved graph with its literature provenance. \revmark{T3-3}The interface supports locating answer-named entities in the retrieved graph, inspecting and editing retrieval state, and reading literature provenance for relationships, while claim-level and scientific judgments remain with the researcher. \system{} consists of two main components: the chat panel on the left and the graph panel on the right (\autoref{fig:final_interface}). \revmark{T4-4}The illustrative scenario in \autoref{sec:case_study} and \autoref{fig:case_study} shows these interactions in sequence.

\vspace*{-0.1cm}
\subsection{Chat Panel}
The chat panel resembles a standard LLM chatbot interface that supports typical conversation through prompts. The chat panel is integrated with the GraphRAG pipeline (\autoref{sec:backend}) with conditional retrieval. \revmark{T4-5}Subquestions that FoodAtlas cannot answer are routed to the standalone LLM path described in \autoref{sec:backend}, and their content appears in the response without graph evidence.

Following the cognitive forcing function (CFF) design, the chatbot first shows an \textit{Answer Context Summary} panel (\autoref{fig:final_interface}-A) before revealing the final answer. The panel describes the retrieved context in a few bullet points. \revmark{T4-5}The summary is generated solely from the retrieved FoodAtlas context before final-response synthesis, previewing the evidence on which the answer will draw. After reviewing the context, the user can click a checkbox to certify that they have ``reviewed the answer context and checked the graph for errors or discrepancies'' and reveal the LLM's answer. In response to the answer-first behavior identified in \inlineDRbox{DR1}, this evidence-first sequence is intended to interrupt answer-first inspection and direct attention to the retrieved context.

To support response--evidence alignment (\inlineDRbox{DR3}), \system{} converts FoodAtlas IDs included in the generated response into hyperlinks. For example, \autoref{fig:final_interface}-A1 mentions ``pineapple'' with entity ID \textit{e1484}. Selecting the hyperlink centers and highlights the corresponding entity in the graph panel, reducing the need to locate an answer-mentioned entity manually within a large retrieved context. Note that this linkage shows where an entity occurs in the retrieved graph, but does not by itself establish that the complete response claim is supported.

\vspace*{-0.1cm}
\subsection{Graph Panel}
The graph panel uses a hybrid hierarchical node-link representation. Cross-type FoodAtlas relationships are shown as node-link edges, while \textit{isA} relationships organize food and chemical entities into navigable groups within type-specific columns. This combination preserves the multi-hop paths needed for verification while allowing users to change the visible level of detail through hierarchical aggregation.

Entities are arranged in three stable columns---\textit{Disease}, \textit{Chemical}, and \textit{Food}---that reflect the typed paths available in FoodAtlas. FoodAtlas links diseases to chemicals and chemicals to foods, but does not directly link diseases to foods. Placing chemicals between the other entity types therefore makes the common disease--chemical--food path spatially explicit and reduces crossings between unrelated relationships.

\vspace*{-0.1cm}
\paragraph{Graph visualization}
At aggregate levels, \system{} bundles edges between visible groups to reduce clutter while preserving the existence of connections between those groups (\inlineDRbox{DR2}). Hovering over or expanding a group exposes the individual entities and relationship types represented by a bundle, such as ``contains'' or ``is positively correlated with.'' A bundle communicates connectivity, not relationship frequency, importance, or certainty.

Entities explicitly mentioned in the generated response are highlighted in yellow. Selecting an entity highlights its incident relationships and opens its Metadata Card. The card contains the entity's FoodAtlas name, literature synonyms, external identifiers, and related entities. For literature-derived relationships, users can expand a relationship to inspect the source identifier and supporting sentence excerpt supplied by FoodAtlas (\autoref{fig:case_study}-B1). These details are disclosed on demand so that provenance inspection does not overwhelm the initial graph overview. \system{} retrieves and presents these records; it neither generates the evidence nor establishes that the source relationship is factually correct.

The graph panel supports zooming and panning, along with a focus+context minimap~\cite{bjork1999focus_context} (\autoref{fig:final_interface}-B). Selecting an entity highlights its incident relationships, and the ``Locate'' controls allow users to move between related entities and return to the selected entity. Together, these interactions support an overview--inspect--return workflow without requiring users to reconstruct their position manually. Buttons directly support expanding or collapsing groups and resetting grouping granularity or the graph panel.

\vspace*{-0.1cm}
\paragraph{Hierarchy construction and controls}
FoodAtlas provides \textit{isA} relationships for food and chemical entities, and \inlineFbox{F3} showed that participants needed to inspect evidence at both broader group and individual-entity levels. We therefore organize these entities using a file-directory-inspired hierarchy. Indentation communicates containment, while familiar folder-style navigation lets users change the visible level of aggregation without losing access to individual entities. This organization preserves the FoodAtlas taxonomy while supporting familiar movement between aggregate groups and individual evidence.

Breadcrumbs show the taxonomic path of each group and allow users to return directly to an ancestor level. When a path contains consecutive groups with only one child, \system{} compresses those levels in the canvas while retaining the complete path in the breadcrumb. This avoids spending display space on uninformative intermediate levels without concealing their position in the taxonomy.

The hierarchy controls support movement between overview and detail, addressing the difficulty reconstructing multi-hop relationships observed in \inlineFbox{F3}. ``Drill Down'' expands a selected group by one taxonomic level so that users can inspect its subgroups or individual entities, whereas ``Merge Up'' restores the parent-level aggregation. When a group contains both direct leaf entities and descendants inside subgroups, expanding it can leave related entities distributed across multiple visible levels. ``Gather'' collects these descendants under the selected group, restoring a coherent aggregate view (\autoref{fig:final_interface}-D).

\vspace*{-0.1cm}
\paragraph{Retrieval inspection and correction}
The discrepancies observed in \inlineFbox{F3} could originate upstream of the displayed graph, including from question expansion and entity mapping. The context control panel therefore exposes the subquestions used for retrieval and the entities and relationships returned for each one. Users can highlight the context associated with a subquestion, replace an entity mapping with another retrieved candidate, delete an irrelevant subquestion, or remove an entity or group from the answer context. Entity and group deletions can be undone before regeneration (\autoref{fig:case_study}-B).

After an edit, \system{} displays the ``Requery With Changes'' control above the minimap (\autoref{fig:final_interface}-B). Activating it reruns the latest request using the revised retrieval state and produces a new context summary and answer. This interaction allows users to examine how an upstream correction affects the downstream response. It does not modify FoodAtlas, validate the edited context, or guarantee that the regenerated answer is correct.

\vspace*{-0.1cm}
\paragraph{Progressive disclosure of graph and domain-specific information}
Displaying structured and domain-aligned information (\inlineDRbox{DR2}) requires balancing access to heterogeneous data and sources against visual complexity. \system{} therefore applies progressive disclosure across the verification workflow. Users begin with a high-level textual summary of the retrieved context, while the graph panel initially presents aggregate groups connected by bundled edges. They can then use the hierarchy controls to reveal individual entities and relationships and open metadata cards to inspect FoodAtlas-supplied provenance. This progression supports inspection from response context to retrieved graph structure and, when needed, to the provenance of individual knowledge-graph relationships.

\vspace*{-0.1cm}
\paragraph{Supported verification tasks}
\revmark{T3-4}For response grounding, the interface supports locating entities named in the answer within the retrieved graph and reading the answer context summary. For retrieval correctness, it supports inspecting each subquestion, its entity mappings with alternative candidates, and the entities and relationships it returned, and editing these before regeneration. For knowledge-graph validity, it supports reading the source identifier and sentence excerpt that FoodAtlas records for a relationship. Judgments of claim-level support and of a relationship's scientific validity remain with the researcher.

\subsection{Design Scope and Alternatives}
The selected representation reflects the verification breakdowns observed in the formative study and the structure of FoodAtlas; we did not experimentally compare it with alternative graph representations. The baseline tables supported inspection of individual triplets but made multi-hop paths difficult to reconstruct, as reported in \inlineFbox{F3}. A completely flat node-link representation would preserve those paths but risk substantial visual clutter for the retrieved contexts encountered in our system. The hybrid representation therefore retains node-link relationships while using the available taxonomy to control aggregation.

\system{} does not encode a single visual measure of verification certainty. The pipeline does not provide a calibrated confidence value shared across generated claims, entity mappings, retrieved relationships, and source quality. For example, an entity-mapping similarity score measures lexical correspondence but does not establish that the mapping or resulting answer is correct. Presenting these heterogeneous signals as a unified certainty encoding could therefore imply unsupported reliability. Instead, \system{} exposes the underlying mappings, relationships, and provenance for inspection. \revmark{T3-5}The design assumes stable identifiers, typed relationships, and a usable \textit{isA} hierarchy, and its three-column layout is specific to FoodAtlas's disease--chemical--food schema; knowledge graphs lacking these properties would need different grouping and layouts, as discussed in \autoref{sec:limitations}.

\subsection{Backend Upgrades}
To support the verification interactions in \system{}, we made various updates to the backend architecture described in~\autoref{sec:backend}. First, to support the hierarchical groupings, we leveraged previously unused \textit{isA} relationships within food and chemical entities to construct their hierarchies. To generate the \textit{Answer Context Summary}, we added a separate invocation that summarizes the retrieved FoodAtlas context before final-response synthesis. Finally, to support context editing and answer regeneration, we modified the pipeline to support regeneration with edits to the answer context.

\section{Illustrative Verification Scenario}
\label{sec:case_study}
Alice is a food-science researcher assessing whether literature-extracted food--chemical--disease relationships are sufficiently reliable for a preliminary review of foods associated with heart disease. She asks \system{}, ``What kinds of food are recommended for patients with heart disease?'' Before revealing the response, she reviews the context summary (\autoref{fig:final_interface}-A), which identifies pineapple and banana as positively correlated with heart disease but contains no foods with chemicals negatively correlated with the disease. Both results conflict with her domain knowledge: these fruits are not typically considered harmful for heart disease, and she expects some beneficial relationships to be available. She therefore treats the discrepancies as signals for further inspection rather than immediately accepting or rejecting the result. After revealing the generated response, she finds that it does not resolve her concerns and turns to the relationships' provenance and the upstream retrieval decisions.

Alice first follows the ``pineapple'' hyperlink in the response to its node in the graph (\autoref{fig:final_interface}-A1; \autoref{fig:case_study}-B). Following the linked entity lets her isolate the graph path supporting the unexpected recommendation and determine whether the concern originates in the generated response or its retrieved evidence. The displayed path indicates that pineapple contains ozone, which is positively correlated with pulmonary heart disease. Although Alice is unfamiliar with the latter association, she knows that ozone is used during food processing but is not typically found naturally within foods, making the \textit{contains} relationship suspect. Rather than dismissing the entire path, she opens pineapple's metadata card and expands the provenance record for that relationship. The source excerpt states that pineapple was ``treated with ozone'' (\autoref{fig:case_study}-B1), suggesting that a treatment was incorrectly extracted as composition. Because FoodAtlas supplies no other source for the relationship, Alice removes pineapple from the answer context.

Alice next investigates why the system retrieved no foods containing chemicals negatively correlated with heart disease. In the context control panel (\autoref{fig:case_study}-C), she locates the relevant subquery, ``What foods are negatively correlated with heart disease?'' Highlighting its results reveals that it returned no entities or relationships. Rather than concluding that FoodAtlas contains no relevant evidence, she inspects the subquery's entity mappings. They show that ``heart disease'' was mapped to ``pulmonary heart disease'' (\autoref{fig:case_study}-D), a narrower condition that may have fewer associated records. From the alternative candidates, she selects the broader entity ``heart diseases.'' She also reviews the other subqueries: one duplicates the relevant question, while two seek foods or chemicals that worsen heart disease and do not serve her inquiry. She removes all three (\autoref{fig:case_study}-C1), retaining only the corrected subquery for regeneration.

Alice then selects ``Requery With Changes'' (\autoref{fig:final_interface}-B). The revised summary and graph contain more foods connected through curcumin and ellagic acid, which are negatively correlated with the corrected ``heart diseases'' entity. She can now trace the retrieved foods through these chemicals to the disease rather than relying on the summary alone. Finding this evidence more relevant, she reveals the regenerated response and checks it against the updated graph. The scenario illustrates how progressive inspection connects an initially questionable response to errors in relationship extraction and entity mapping, allowing Alice to revise the retrieved context and reassess the response.

\section{Qualitative System-Use Study}
\label{sec:user_study}
\revmark{T2-6}We conducted a qualitative system-use study, supplemented by descriptive questionnaire responses, to examine how food-science researchers used \system{} to inspect generated GraphRAG responses and retrieved graph evidence. The study characterizes observed interactions and reasoning strategies rather than comparing interface performance or estimating population-level effects.

\subsection{Study Design}
To operationalize \textbf{RQ3}, the study considers two subquestions: \textbf{RQ3a:} \textit{How do participants inspect and relate generated responses and retrieved graph evidence using \system{}?} \textbf{RQ3b:} \textit{How does participants' food-science knowledge shape their verification strategies and recognition of knowledge-graph limitations?}
We adopted the same quiz structure, think-aloud protocol, and post-task questionnaires as in the formative study. The three quizzes asked participants to assess claims about spinach and gout, honey and diabetes, and foods relevant to heart disease, and to compose an answer using information obtained through the interface. Unlike the formative study, the controlled error-injection mode described in~\autoref{par:error_injection} was not activated. The live pipeline and FoodAtlas could nevertheless produce discrepancies during a session. The evaluated implementation used GPT-4o for its LLM invocations. Accordingly, the findings refer only to the outputs and interactions observed in these six sessions.

\paragraph{Participants}
\revmark{T2-7}We recruited five food-science graduate students and one senior food-science researcher through institutional and external research networks.
\revmark{T4-6}We refer to them as P1--P6; these identifiers are independent of those used for the formative-study participants.
Their median age was 26.5 years, and two identified as male and four as female. All had at least three years of food-science experience, with a median of seven years. On 7-point familiarity scales, where higher values indicate greater familiarity, median ratings were 6 for food-science concepts, 5 for LLM-based services, and 4.5 for knowledge graphs. Participant-level demographic and familiarity values are provided in the supplemental materials.

\paragraph{Procedure}
The system-use study protocol was approved by the institutional review board.
Each remote session was conducted via Zoom and lasted approximately 60 minutes.
After an initial study overview and verbal consent, participants completed a pre-questionnaire on their background and experience.
Participants then completed a short practice session to familiarize themselves with the interface before proceeding to the main tasks.
As in the formative study, participants completed three quizzes via a Google Form. 
For each quiz, they interacted with the system while thinking aloud, submitted their answer, and completed a post-questionnaire.
With participants' consent, we recorded screen activity and audio through Zoom after the practice session.
Finally, participants took part in a semi-structured interview focusing on their verification strategies and interaction experience. 
Participants received \$20 USD as compensation.

\paragraph{Analysis}
\revmark{T4-7}One author conducted this descriptive qualitative analysis by reviewing the session recordings alongside contemporaneous observation notes and marking and grouping episodes inductively without a predefined codebook. We marked episodes where participants inspected evidence, compared panels, traced retrieval or graph discrepancies, edited context, or invoked domain knowledge. We compared episodes across participants and organized patterns around RQ3a and RQ3b. We report participant-attributed actions and quotations as evidence without using their frequency to estimate prevalence. Questionnaire responses provide descriptive context and are not treated as evidence of comparative effectiveness.

\subsection{Findings}
Across the six sessions, the findings show mixed experiences with the evidence-first reveal and cross-layer discrepancy tracing (\textbf{RQ3a}), as well as how domain knowledge guided verification and exposed knowledge-graph limitations (\textbf{RQ3b}).

\paragraph{Evidence-first interaction produced varied experiences (\textbf{RQ3a})}
Participants sometimes inspected retrieved evidence before revealing the response. P3 reported that they ``were looking for more answers on the graphical side before checking the checkbox.'' The interaction did not produce a uniform workflow: P2 found the delayed reveal confusing because the initial summary felt ``not directly related to the prompt task,'' and P2 and P4 occasionally overlooked the checkbox. Thus, the sequence prompted evidence inspection in some sessions but introduced interpretation and discoverability costs in others.

\paragraph{Tracing discrepancies across response, graph, and retrieval (\textbf{RQ3a})}
Participants compared entities in generated responses with the retrieved graph and investigated where the panels diverged. P2 found that \textit{honey} appeared in the response but not in the graph and attributed it to the model's prior knowledge. This comparison identified content without visible support in the retrieved graph, although it did not establish the content's source or factual correctness. At the retrieval layer, P6 traced sparse heart-disease results to a mapping from ``heart disease'' to ``pulmonary heart disease.'' After selecting ``heart diseases'' and requerying, P6 received additional context that ``make[s] so much more sense now.'' Participants also assigned different roles to the panels: P5 described responses as more general and the graph as more specific but potentially incomplete. Consequently, missing graph evidence did not automatically lead participants to reject response content. Together, these episodes show how participants traced discrepancies across generated content, retrieved evidence, and mapping decisions.

\paragraph{Domain knowledge guided verification and exposed evidence limitations (\textbf{RQ3b})}
Participants used prior knowledge to formulate questions, choose inspection paths, and judge when graph evidence was insufficient. P6 traced paths from foods to chemicals and then to diseases, explaining, ``I kind of knew the answer already, but following the links helped me notice unexpected relationships in the graph.'' P5 decomposed quiz questions into narrower prompts and inspected their graph contexts separately. These strategies show how prior knowledge could prompt acceptance, further querying, or skepticism toward the same retrieved evidence.

Domain knowledge also helped participants recognize limitations in that evidence. P5 and P6 questioned the small number of foods connected to heart disease and interpreted the coverage as incomplete. They also noted that dosage, intake, and concentration information needed to judge relationships was missing. P6 investigated the improbable claim that blueberries contain curcumin and found that the source instead described curcumin being injected into blueberries as an experimental treatment. This provenance check exposed a possible extraction error in which an experimental treatment was represented as a natural component. Across these episodes, domain knowledge directed inspection while revealing incomplete coverage, missing qualifications, and a potentially incorrect relationship.

\paragraph{Descriptive questionnaire context}
Questionnaire responses were broadly consistent with these observations. \revmark{T4-8}The post-task questionnaire included six items rated on a 1--7 agreement scale, with higher values indicating stronger agreement: distinguishing graph-grounded from generated content, attempting verification, comfort using the system without checking, feeling responsible for deciding correctness, confidence in the answer, and ability to check the answer; the full wording appears in the supplemental materials. Across the 18 task-level responses, the median was 5.5 for attempting verification and 5 for perceived ability to verify, distinguish graph-supported from generated content, judge correctness, and provide a confident answer. Comfort using the system without checking was more mixed (median: 4). These repeated ordinal responses are not evidence of population-level or causal effects.

\section{Discussion}

\subsection{Informative friction}
We found that informative friction directed some participants’ attention to retrieved evidence before the generated answer, while introducing costs to understandability and discoverability.
In the user study, P3 inspected the graph before revealing the response, showing how the intervening information could contribute to the task rather than serving only as an interruption. The interaction was not uniformly successful: P2 found the sequence confusing, and P2 and P4 sometimes overlooked the reveal control.

These observations expose a tension between directing attention to evidence and keeping the interaction understandable. Informative friction should pair delay with immediately inspectable, task-relevant evidence while clarifying the required action and the relationship between preview and response. \revmark{T5-1}An LLM invocation produces the context summary but may still omit or distort retrieved information, so the summary could become an anchor for verification. Readers who treat the summary as a faithful digest may inspect the graph less closely, allowing omissions or distortions in the summary to shape what they verify and leaving verification dependent on the LLM’s selection of evidence. While we did not evaluate how users anchor on the summary, we treat this as a promising direction for comparative and longitudinal studies of effects on reliance, usability, and habituation.

\subsection{Progressive verification for human oversight}
Recent studies call for human oversight in LLM-based systems, but naively adding verification support might transfer the load to users~\cite{gaube2026keeping_eye, passi2025agentic_oversight, dhanorkar2026human_oversight, naik2026catch22}. Because more information does not ensure effective oversight and extended AI use may weaken relevant skills, verification effort and fatigue remain design concerns~\cite{rismani2026use_oversight, mitchell2026out_loop, crescitelli2026verification_cost, fan2026verification_load}.
Discrepancies such as P6’s entity mapping trace show how GraphRAG distributes verification across layers, while participants’ varied paths motivate multiple inspection routes.
Progressive verification addresses this burden by staging inspection from an overview to detail only where discrepancies or doubt warrant closer examination.

\system{} operationalizes progressive verification by externalizing intermediate evidence for users to inspect and repair across layers. Before the answer is revealed, users see the context summary and retrieved evidence graph, which gives a broad view of what the answer will rest on. The intended order is the context summary, then aggregate groups, then provenance for selected relationships, so detail appears only where doubt or discrepancies warrant it. When a group looks incomplete or a relationship looks doubtful, the user expands the hierarchical group to individual relationships and opens their source identifiers and literature excerpts. Answer hyperlinks connect named entities to graph nodes, while the context control panel supports entity remapping or retrieval state edits before regeneration closes the loop. VArify is intended for deliberate verification sessions, such as a food-science researcher auditing literature-derived relationships before relying on them, and the interface’s inspection depth is sized for this use.
Still, repair and progressive inspection leave two questions open: how user edits shape regeneration and how the added inspection targets affect verification effort. Incorrect edits to subquestions, entities, or mappings can propagate into regenerated answers and reinforce prior expectations, and regenerated answers are not validated against the edit. The second question remains unresolved, leaving open whether those targets reduce or redistribute verification effort.

\subsection{Verification boundaries and unknown unknowns}
Verification depends on what the system retrieves and what the user knows. \autoref{fig:verification_boundary} crosses these dimensions: the horizontal axis indicates whether information is retrieved, and the vertical axis indicates whether it is familiar to the user. The system provides a concrete basis for action in three of the quadrants. In Aligned or conflicting context, the user can compare retrieved relationships against prior knowledge, allowing the evidence to confirm or contradict what is known. In Missing context, the user can examine subquestions and entity mappings, allowing an expected item’s absence to be traced to retrieval or mapping. In New or irrelevant context, the user can inspect relationships and expand their FoodAtlas or literature sources, while relevance and correctness remain dependent on the sources and the user’s judgment.

For potentially relevant information that is neither familiar nor retrieved, i.e., unknown unknowns, the system provides no visible cue. Such unknown unknowns fall outside \system{}'s observable boundary and may remain undetected without additional evidence or expertise. Making this boundary explicit helps researchers interpret a verified answer as supported within an observable scope whose limits remain consequential.

\begin{figure}[t]
    \centering
    \resizebox{0.8\columnwidth}{!}{%
    \begin{tikzpicture}[font=\sffamily, every node/.style={align=center}]
        \def\W{4.2}
        \def\H{2.25}

        \fill[gray!18]   (0,\H) rectangle (\W,2*\H);
        \fill[orange!20] (\W,\H) rectangle (2*\W,2*\H);
        \fill[violet!15] (0,0) rectangle (\W,\H);
        \fill[teal!20]   (\W,0) rectangle (2*\W,\H);

        \draw[thick] (0,0) rectangle (2*\W,2*\H);
        \draw[thick] (\W,0) -- (\W,2*\H);
        \draw[thick] (0,\H) -- (2*\W,\H);

        \node[font=\large\bfseries] at (0.5*\W,1.55*\H) {Unknown unknown};
        \node[font=\small, black!65] at (0.5*\W,1.30*\H) {Outside observable boundary};

        \node[font=\large\bfseries] at (1.5*\W,1.58*\H) {New or irrelevant\\context};
        \node[font=\small, black!65] at (1.5*\W,1.27*\H) {Inspect relationships and sources};

        \node[font=\large\bfseries] at (0.5*\W,0.58*\H) {Missing context};
        \node[font=\small, black!65] at (0.5*\W,0.31*\H) {Inspect retrieval and mappings};

        \node[font=\large\bfseries] at (1.5*\W,0.60*\H) {Aligned or conflicting\\context};
        \node[font=\small, black!65] at (1.5*\W,0.27*\H) {Compare evidence and knowledge};

        \node[font=\large\bfseries] at (0.5*\W,2*\H+0.42) {Not retrieved};
        \node[font=\large\bfseries] at (1.5*\W,2*\H+0.42) {Retrieved};
        \draw[->, thick] (0,2*\H+0.82) -- (2*\W,2*\H+0.82)
            node[midway, above, font=\large\bfseries] {Information in retrieved context};

        \node[font=\large\bfseries, rotate=90] at (-0.48,1.5*\H) {Unknown};
        \node[font=\large\bfseries, rotate=90] at (-0.48,0.5*\H) {Known};
        \draw[->, thick] (-0.90,2*\H) -- (-0.90,0)
            node[midway, above, rotate=90, font=\large\bfseries] {User knowledge};
    \end{tikzpicture}%
    }
    \caption{Conceptual organizer for the verification boundary. Three quadrants provide an observable signal because information is either familiar to the user or present in the retrieved context, providing a starting point for inspection through \system{}. Potentially relevant information unfamiliar to the user and absent from retrieval remains outside this boundary.}
    \label{fig:verification_boundary}
\end{figure}

\subsection{Limitations and Future Work}
\label{sec:limitations}
\textit{System.} The system limitations concern the graph properties that the design depends on and its transfer to other knowledge graphs. \revmark{T3-6}The workflow requires inspectable retrieval decisions, editable context, and provenance, while the representation assumes stable identifiers, typed relationships, and useful grouping information. FoodAtlas's three-column layout fits its disease--chemical--food schema; graphs without hierarchies or with multi-parent need other grouping strategies and layouts. \revmark{T4-10}Its analytical scope also excludes graph-analytic question types beyond FoodAtlas's relationship queries.

\textit{Studies.} The study limitations define the evidentiary reach of the results: small samples and short sessions suit requirements elicitation and qualitative characterization. The injected errors and advance warnings in the formative study may have heightened attention to discrepancies. \revmark{T4-10}Because both qualitative analyses relied on a single coder without a predefined codebook, the patterns reflect one analyst's reading, anchored to quoted actions and participant-attributed episodes. \revmark{T5-3}The studies did not measure task time, interaction counts, workload, or entity-mapping recall or precision, leaving that retrieval stage's coverage and accuracy for future evaluation.

\textit{Claims.} \revmark{T4-10}The claims are limited to descriptive accounts supported by quoted actions; consequently, they neither establish improved verification or reduced overreliance nor estimate the frequency of observed behaviors. \revmark{T4-10}A controlled comparison with planted errors and outcome measures is the appropriate next study of whether this design improves verification.

Future work could examine user- or query-adaptive grouping~\cite{Qiu2025vadis, sarthi2024raptor, wang2025speculative_rag} and layer-specific uncertainty cues~\cite{padilla2021uncertainty_visualization, beigi2024rethinkinguncertainty, passi2025overreliance, vasconcelos2025generation_probabilities}. Provenance-identified errors could support rule-based or few-shot correction~\cite{Ott2023rule_based_KG, mou2024fewshot_KG}.
\section{Conclusion}
We investigated the verification of GraphRAG responses at the response, retrieval, and knowledge-graph layers. \revmark{T1-7}A formative study of verification breakdowns in a baseline interface yielded design requirements for \system{}. \system{} visually presents evidence before answers and supports food-science researchers' coordinated inspection and correction of answers, retrieval state, and provenance-linked retrieved knowledge subgraphs. A system-use study documented inspection before answer reveal, discrepancy tracing, and recognition of knowledge-graph limitations. These observations do not establish reduced overreliance or comparative effectiveness, but they motivate interfaces that expose evidence and the limits of verifiability.
\bibliographystyle{abbrv-doi-hyperref}

\balance
\bibliography{references}








\end{document}